\documentclass[12pt, draftclsnofoot, onecolumn]{IEEEtran} 
\ifCLASSINFOpdf
\else
\fi

\usepackage{cite}
\usepackage{stfloats}
\ifCLASSINFOpdf
   \usepackage[pdftex]{graphicx}
   \graphicspath{{../pdf/}{../jpeg/}}
   \DeclareGraphicsExtensions{.pdf,.jpeg,.png}
\else
   \usepackage[dvips]{graphicx}
   \graphicspath{{../eps/}}
   \DeclareGraphicsExtensions{.eps}
\fi

\usepackage[cmex10]{amsmath}
\usepackage{mathrsfs,color,amssymb}
\usepackage{amsfonts}

\usepackage{algorithm}
\usepackage{algorithmic}
\allowdisplaybreaks
\usepackage{hyperref}
\hypersetup{hidelinks}

\ifCLASSOPTIONcompsoc
\else
  \usepackage[caption=false,font=footnotesize]{subfig}
\fi

\usepackage{hyperref}
\usepackage{graphicx}
\usepackage{epstopdf}

\begin{document}

%

\title{Signal Processing for MIMO-NOMA:\\
Present and Future Challenges}

%
%
%

\author{Yongming~Huang,~\IEEEmembership{Senior Member,~IEEE,}
        Cheng~Zhang,~\IEEEmembership{Student Member,~IEEE,}
        Jiaheng~Wang,~\IEEEmembership{Senior Member,~IEEE,}
        Yindi~Jing,~\IEEEmembership{Member,~IEEE,}
        Luxi~Yang,~\IEEEmembership{Member,~IEEE,}
        and Xiaohu~You,~\IEEEmembership{Fellow,~IEEE}
\thanks{
This work has been accepted by the IEEE Wireless Communications, the special issue of non-orthogonal multiple access for 5G. Copyright may be transferred without notice, after which this version may no longer be accessible.

Y. Huang (Corresponding Author), C. Zhang, J. Wang, L. Yang, and X. You are with the National Mobile Communications Research Laboratory, the School of Information Science and Engineering, Southeast University, Nanjing 210096, P. R. China (email: {huangym, zhangcheng1988, jhwang, lxyang, xhyu}@seu.edu.cn).

Y. Jing is with the Department of Electrical and Computer Engineering, University of Alberta, Edmonton, Canada, T6G 1H9 (email: yindi@ualberta.ca).
}}

\maketitle

\begin{abstract}
Non-orthogonal multiple access (NOMA), as the newest member of the multiple access family, is envisioned to be an essential component of 5G mobile networks. The combination of NOMA and multi-antenna multi-input multi-output (MIMO) technologies exhibits a significant potential in improving spectral efficiency and providing better wireless services to more users. In this article, we introduce the basic concepts of MIMO-NOMA and summarize the key technical problems in MIMO-NOMA systems. Then, we explore the problem formulation, beamforming, user clustering, and power allocation of single/multi-cluster MIMO-NOMA in the literature along with their limitations. Furthermore, we point out an important issue of the stability of successive interference cancellation (SIC) that arises using achievable rates as performance metrics in practical NOMA/MIMO-NOMA systems. Finally, we discuss incorporating NOMA with massive/millimeter wave MIMO, and identify the main challenges and possible future research directions in this area.
\end{abstract}

%
\IEEEpeerreviewmaketitle

\section{Introduction}
As a promising multiple access (MA) technique for the fifth generation (5G) communications, non-orthogonal MA (NOMA) can significantly improve the spectral efficiency and user fairness of mobile communication networks \cite{Dai_mag}. For orthogonal MA (OMA), a main issue is the low spectral efficiency when the wireless (time/frequency/code) resources are allocated to users with poor channel conditions. NOMA alleviates this problem by using superposition coding (SC) and successive interference cancellation (SIC) and enabling users with significantly different channel conditions to share the same resource block. Thus, NOMA is regarded as one of the enabling technologies to meet the requirements of the high rate, dense coverage, massive connectivity and low latency in 5G. In NOMA, a good balance between the spectral efficiency and the user fairness can be achieved via proper user pairing and power allocation policies, with affordable costs in SIC complexity and signaling overhead.

Besides NOMA, multi-input and multi-output (MIMO) is another popular technique to increase the spectral efficiency by exploiting resources in the spatial dimension.
By assuming perfect channel state information (CSI) at the transmitter, it has been shown that the capacity region of a multiuser MIMO downlink channel can be achieved via dirty-paper coding (DPC).
However, due to the difficulty of implementing DPC in practice, the combination of more practical MIMO transceivers with NOMA is preferable, which approaches the performance of DPC under certain channel conditions \cite{Zhiyong_quasi}. Further, when the transmit antennas are less than the total receive antennas, DPC itself becomes non-optimal and in this case the additional power dimension offered by NOMA is necessary.

The main goal of this work is to investigate signal processing techniques for MIMO-NOMA and explore potential research directions on this topic. Specifically, the basic concepts for MIMO-NOMA are elaborated, followed by an introduction of the performance criteria and design problems in MIMO-NOMA. Then, we review the existing works on MIMO-NOMA along two main routes: single-cluster and multi-cluster MIMO-NOMA. In the former case all users except the weakest one conduct SIC, while in the latter case users are first partitioned into different clusters. In MIMO-NOMA, beamforming can be applied along with intra-cluster SIC to suppress both intra-cluster interference and inter-cluster interference. The full benefit of MIMO-NOMA is achieved by a jointly design of user clustering, beamforming, power allocation, and SIC.

In existing NOMA/MIMO-NOMA works, it is often assumed that decoding in SIC is perfect without errors. Hence, achievable rates have been widely used in NOMA system designs. In this paper, we would like to point out an important issue, called the SIC-stability \cite{Jianyue_Optimal}, that arises from using achievable rates as performance metrics in practical NOMA systems. The optimal NOMA strategy based on achievable rates may cause an adverse effect to SIC, e.g., equal power allocation that leads to the largest error propagation \cite{Wang_SPL}. We demonstrate such an effect on SIC through simulations, where a tradeoff is observed between the accuracy of decoding the weak user's message and that of the stronger user's message.

Finally, we discuss the incorporation of NOMA and massive/millimeter wave (mmWave) MIMO, which are 5G enabling technologies as well, as the future research directions. The main challenges result from the massive antenna configuration and the large path loss, which make traditional linear channel estimation methods ineffective and the popular assumption in MIMO-NOMA, e.g., perfect instantaneous CSI at the BS, impractical. Possible solutions and future directions are provided from three aspects: 1) channel estimation and limited feedback, 2) hybrid massive MIMO-NOMA, and 3) statistical massive MIMO-NOMA.

\section{Basic Concepts and Problems in MIMO-NOMA}
In this section, we first introduce the essential principle of NOMA via a simple single-input single-output (SISO) scenario. Then, the concepts of MIMO-NOMA are introduced and the performance criteria and design problems are elaborated.
\subsection{Principle of NOMA}
To exhibit the essential principle of NOMA, consider a two-user SISO-NOMA system as an example, where
the BS serves the two users, a far user $U_1$ and a near user $U_2$, sharing the same resource block
with different transmit powers $P_1 > P_2$.
At the user side, $U_1$ decodes its message by treating $U_2$'s message as noise, while $U_2$ first decodes the message of $U_1$ and then decodes its own message after removing $U_1$'s message. In this way, both users have full access to the whole resource block, and the near user can decode its message without interference from the far user. Ideally, by using NOMA, the achievable rate of $U_1$ and $U_2$ are given respectively by
\begin{equation}\label{siso}
R_1=\log_2\left(1+\frac{P_1|h_1|^2}{1+P_2|h_1|^2}\right),\quad R_2=\log_2(1+{P_2|h_2|^2}),
\end{equation}
where $h_i$ is the channel coefficient of $U_i$. In the case of $K$ users, given a user order $U_1, \cdots, U_K$, SIC is applied such that $U_i$ decodes the messages of $U_1,U_2,\cdots,U_i$ sequentially and while decoding $U_j$'s message ($j\le i$), the interference from $U_1,U_2,\cdots,U_{j-1}$ is removed based on previous decoding results and the interference from $U_{j+1},\cdots,U_K$ is treated as noise.

\subsection{From SISO-NOMA to MIMO-NOMA}
In MIMO-NOMA, the BS is equipped with multiple antennas and the users may have a single or multiple antennas. Similar to SISO-NOMA, SIC is implemented at the user side to suppress interference.
While in SISO-NOMA the main focus is on optimizing power allocation among users, given the extra spatial degrees of freedom in MIMO-NOMA, it is possible to eliminate the user interference via beamforming in both the power and spatial domains. This, however, results in more complicated MIMO-NOMA design problems. For example, in SISO-NOMA, the SIC order usually depends directly on channel gains, e.g., a user with smaller channel gain has a higher SIC order (i.e., decoded first). However, in MIMO-NOMA, the effective channel gains are related to particular beamforming designs, which makes the beamforming and SIC ordering optimization coupled.

\begin{figure}[t]
\centering
\includegraphics[scale=0.9]{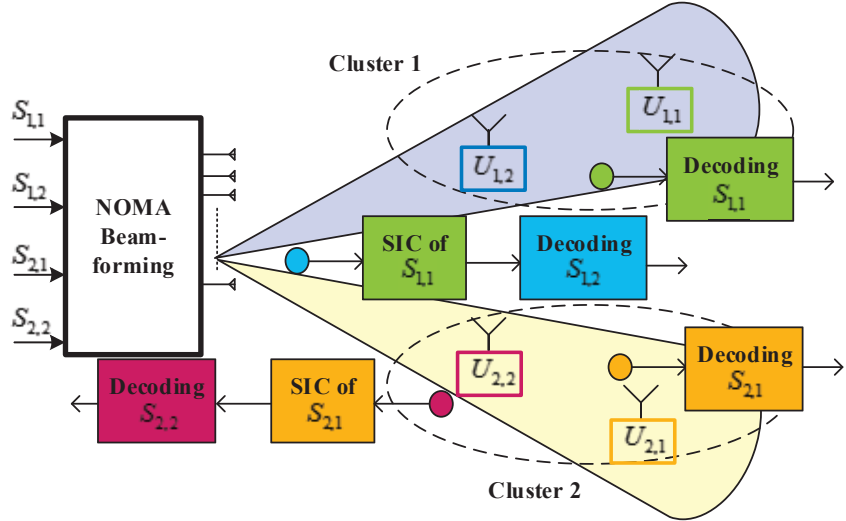}
\captionsetup{margin=5pt,font=small}
\caption{A MIMO-NOMA system with $N=1$, $C=2$, $\bar K=2$.}\label{MIMO-NOMA}
\end{figure}

In the following, we introduce more details about MIMO-NOMA via a general multi-cluster MIMO-NOMA downlink system, which consists of a BS with $M$ antennas and $K$ users each with $N$ antennas.
Assume that the users are partitioned into $C$ clusters with $\bar K$ users in each cluster. Denote the users in cluster $c$ by $U_{c,1},\cdots,U_{c,\bar K}$, and their transmitted messages by $S_{c,1},\cdots,S_{c,\bar K}$, respectively. For simplicity, we assume that the SIC order in cluster $c$ is $U_{c,1},\cdots,U_{c,\bar K}$. Generally, $U_{c,1}$ with the highest SIC order is also called the weakest user, while $U_{c,{\bar K}}$ is called the strongest user. An example of a MIMO-NOMA system with $C=2$, $\bar K=2$, $N=1$ is shown in Fig. \ref{MIMO-NOMA}. To suppress intra-cluster interference, SIC is conducted within each cluster. Beamforming is conducted among different clusters to suppress inter-cluster interference, which is not possible for SISO-NOMA since no extra spatial degree of freedom is available. For sufficient BS antennas, beamforming can also be utilized to suppress or eliminate intra-cluster interference in each cluster.

Similar to SISO-NOMA, in MIMO-NOMA, the most widely used performance metric is the \textit{achievable rate} \cite{Minorization_Hanif}. Specifically, given an SIC order $U_{c,1},\cdots,U_{c,\bar K}$, to guarantee a successful decoding of message $S_{c,m}$ at users $U_{c,m},\cdots,U_{c,\bar K}$, the achievable rate of $U_{c,m}$ is determined by the minimum signal-to-interference-plus-noise ratio (SINR) of its message among all these decodings, which is given by
\begin{equation}\label{achievable rate}
R_{c,m}=\min\limits_{j}\{\log_2(1+{\rm SINR}_{c,j}^{c,m}),j\ge m\},
\end{equation}
where ${\rm SINR}_{c,j}^{c,m}$ denotes the SINR level for decoding $S_{c,m}$ at $U_{c,j}$.
Another two important performance metrics are \textit{outage probability}\cite{Ding_Massive} and \textit{symbol error rate (SER)}. The major design problems in MIMO-NOMA are optimization of the beamforming, power allocation, user clustering, and SIC ordering, either jointly or partially, under some performance metric.

\section{Overview of Existing Works}\label{overview}
In this section, we review some representative works on MIMO-NOMA and discuss their limitations.
The literature is classified into single-cluster and multi-cluster MIMO-NOMA. Unless otherwise mentioned, the reviewed works are based on perfect instantaneous CSI at the BS.

\subsection{Single-Cluster MIMO-NOMA}
For single-cluster MIMO-NOMA, all $K$ users share the same resource block and the cluster index $c$ can be omitted.

In \cite{Minorization_Hanif}, M. F. Hanif et al. tried to find the beamforming and power allocation that maximize the sum-rate of single-antenna users with a given SIC order. Two constraints were considered: 1) total transmit power constraint and 2) the effective power of $S_i$ should be no less than that of $S_j$ for $i<j$ at any user side. The latter one is actually an implicit constraint to protect the weak users' performance. The resulting problem is, however, nonconvex, and thus an iterative algorithm based on successive convex approximation was proposed. In particular, if all users' channels have the same direction, then the SIC order is determined by the channel norms and the achievable rate of each user reduces to $R_{k}=\log_2(1+{\rm SINR}_{k}^{k})$, which simplifies the optimization problem to some extent. Given the suboptimal solution in \cite{Minorization_Hanif}, it is unknown if there is a gap between the proposed MIMO-NOMA design and the optimal DPC based MIMO transmission. The effect of users' channel condition, e.g., channel norm and channel correlation, on the performance is to be investigated, especially for users with similar distances to the BS.

In \cite{Zhiyong_quasi}, Z. Chen et al. first introduced the concept of the quasi-degradation for two single-antenna users, i.e., the situation where the ratio of two users' channel norms is no less than the value of a decreasing function of the angle between the channels. In this case, there exists a closed-form beamforming design that minimizes the transmit power of the NOMA system under SINR constraints and achieves the similar performance to that of DPC. For the user cluster with non-quasi-degradation channels, zero-forcing beamforming (ZFBF) is preferable especially for a large channel angle. Consequently, a hybrid NOMA (H-NOMA) via dynamically switching between the NOMA beamforming and ZFBF can be implemented within each user cluster. Unfortunately, so far the optimal H-NOMA beamformer is only known in the two-user case but unknown in the general $K$-user ($K > 2$) case.


\subsection{Multi-Cluster MIMO-NOMA}
Packing a large number of users in one cluster is not beneficial to exploit spatial degrees of freedom in MIMO-NOMA and also increases the complexity of the beamforming design and SIC. Therefore, multi-cluster MIMO-NOMA ($C>1$) that incorporates space division multiple access (SDMA) with NOMA was proposed \cite{Kim_Non,Kim_Design,Ali_Non,Choi_Minimum,zhiyong_beamforming}. Nevertheless, in this case inter-cluster interference further complicates the beamforming design.

\subsubsection{Insufficient BS Antennas}
When the total number of the receiver antennas of the users in the same resource block is larger than the antenna number at the BS, the spatial degrees of freedom are insufficient for complete SDMA transmission, so in this case NOMA is naturally introduced due to its additional power dimension.


In \cite{Kim_Non} and \cite{Kim_Design}, B. Kim et al. and J. Kim et al. considered a MIMO-NOMA system with $C=M, \bar K=2, N=1$ (i.e., $M$ clusters, two users in each cluster, and single-antenna users) and proposed two sub-optimal transmission schemes using equal power allocation among clusters. Due to the limited antennas at the BS, two users in each cluster share the same beamformer. In \cite{Kim_Non}, to eliminate inter-cluster interference, ZFBF was adopted based on the user's channel with a larger norm (termed as the strong user) in each cluster. Then, user clustering is devised to make the two users within each cluster to have a strong channel correlation and a large difference in channel norms, which result in low inter-cluster interference and high NOMA performance. Based on the designed beamforming and user clustering, intra-cluster power allocation was further optimized under the weak user's rate constraint \cite{Kim_Non}. In \cite{Kim_Design}, the fractional transmit power control (FTPC) was used for intra-cluster power allocation to match the SIC order. Then with the similar heuristic user clustering approach as in \cite{Kim_Non}, a majorization minimization (MM) approach was used to optimize the beamformer.
Note that in \cite{Kim_Non} and \cite{Kim_Design} the beamforming, user clustering, and power allocation were separately determined but not jointly optimized.

In \cite{Ali_Non}, M. S. Ali et al. proposed a new beamforming design for the MIMO-NOMA system with $C=M, N=1$ (i.e., $M$ clusters, single-antenna users). While using the similar user clustering as in \cite{Kim_Non,Kim_Design}, \cite{Ali_Non} proposed to use the dominant right-singular vector of each cluster channel matrix as the cluster representative channel, and developed a joint ZFBF and decoding scaling approach to reduce inter-cluster interference. Noticing the dominant effect of strong users on the sum-rate performance, the users with the largest channel norms are selected as the strongest users in each cluster. Then, each cluster is allocated with the power proportional to the user number in the cluster. Meanwhile, intra-cluster power allocation is optimized to maximize the sum-rate under two constraints:  1) users' minimum rate requirements and 2) a sufficient difference between the power of the desirable message and the total power of the messages with lower SIC order for SIC reliability. The scheme proposed in \cite{Ali_Non} depends on the assumption of distinguishable near-far-user effects, i.e., a sufficient condition for the proposed beamforming and decoding scaling design to be efficient, which, however, may not hold for randomly deployed users.


\subsubsection{Sufficient BS Antennas}
Although NOMA may be more necessary for systems with insufficient BS antennas, it can also be used in systems with a large number of antennas (larger than the total number of all users' antennas).

In \cite{Choi_Minimum}, J. Choi proposed a two-stage beamforming scheme for the multi-cluster MIMO-NOMA system with $N=1, \bar K=2$ (i.e., two users in each cluster, and single-antenna users), to minimize the transmit power with users' SINR constraints.
At the first stage, the singular value decomposition (SVD)-based beamforming is used. Given sufficient BS antennas, \textit{multiple} beams can be allocated to each cluster. This is different from the ZFBF-like inter-cluster beamforming in \cite{Kim_Non, Kim_Design,Ali_Non}, where only a single beam is available for each cluster. At the second stage, the intra-cluster beamforming is optimized over the effective channel in each cluster, but the obtained result is not guaranteed to be optimal. Then, based a good near-far user grouping, a greedy user clustering approach was proposed by exploiting the channel correlation or angle.

In \cite{zhiyong_beamforming},  Z. Chen et al. extended the H-NOMA scheme in \cite{Zhiyong_quasi} to the multi-cluster case, where each cluster has two users. The closed-form beamformers were proposed without the condition on near-far-user distribution.
The so-called PH-NOMA scheme also has two stages: the inter-cluster beamforming at the first stage is based on an orthogonal projection, which is similar to the SVD-based one in \cite{Choi_Minimum}; at the second stage, the closed-form beamformer in \cite{zhiyong_beamforming} or ZFBF is adopted for each cluster based on the quasi-degradation condition. Then, the principle of the user clustering is to form more user pairs satisfying the quasi-degradation condition. So far, PH-NOMA is only applicable for two users in each cluster, and its generalization to the $\bar K$-user clustering is still absent.


Note that, in this section, we only consider narrow-band MIMO-NOMA systems with the focus on the joint design in the spatial and power domains. So far, there are few works on
wide-band MIMO-NOMA. Nevertheless, the extension of wide-band NOMA \cite{Jianyue_Optimal} to MIMO is worth studying as well due to its importance in practical wide-band 5G systems.
\section{Future Directions}
In this section, we first point out an important issue, called the SIC-stability, that arises from using achievable rates as performance metrics in NOMA or MIMO-NOMA systems. 
Then, we discuss key challenges of incorporating NOMA in massive/millimeter-wave MIMO, based on which several future directions are proposed.
\subsection{SIC-Stability}
Most existing NOMA/MIMO-NOMA works adopt achievable rates as performance metrics. Ideally, at any SIC step, if the corresponding transmission rate is lower than the achievable rate shown in \eqref{achievable rate}, SIC is often assumed to be successful with perfect decoding. However, for systems with practical modulation and finite coding length, decoding errors are inevitable, which cause error propagation in SIC and lead to remarkable performance degradation. A well designed NOMA scheme shall avoid error propagation and maintain the SIC-stability \cite{Jianyue_Optimal}. However, using achievable rates may play an opposite role in NOMA designs. For example, consider the two-user SISO-NOMA system as illustrated in Section II-A. With the aim to maximize the sum rate of the two users, the optimal transmit strategy derived from \eqref{siso} may result in equal power allocation \cite{Wang_SPL}, which is the worst situation for SIC and leads to the largest error propagation.

Some attempts on improving the SIC-stability can be seen in \cite{Kim_Design,Ali_Non,Shanzhi_Pattern} . For MIMO-NOMA, SIC-residue due to imperfect decoding was embodied in the achievable rate of the strong user \cite{Kim_Design}. In \cite{Ali_Non}, minimum received power differences among different users were set as constraints in the transmission optimization for reliable SIC.
For SISO-NOMA, conditions were identified to avoid equal power allocation under different criteria \cite{Jianyue_Optimal,Wang_SPL}. A new NOMA scheme based on code pattern, called pattern division multiple access (PDMA), was proposed in \cite{Shanzhi_Pattern} where PDMA patterns are designed to offer different order of transmission diversity to multiple users and thus alleviate error propagation in SIC.

\begin{figure}[t]
\centering
\includegraphics[scale=0.6]{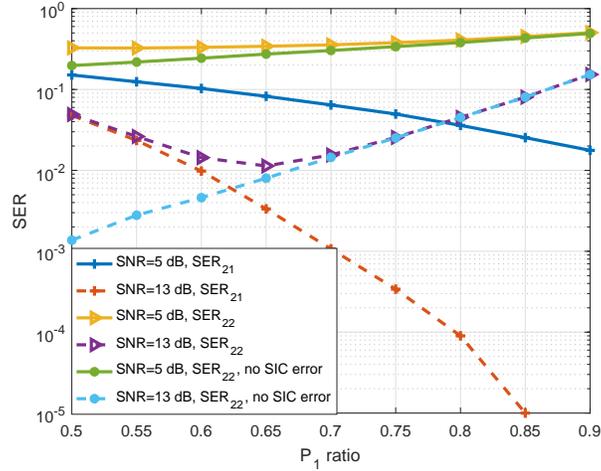}
\captionsetup{margin=5pt,font=small}
\caption{The effect of power allocation on the SER of each decoding in the strong user side. The scalar channel and the receiver noise power are normalized to $1$. $P_1$ ratio denotes the one of $P_1$ to the total power.}\label{power_ser}
\end{figure}
All these attempts are based on the understanding that allocating more power to weak users can alleviate error propagation in SIC and thus is favorable for SIC-stability. However, doing so may increase the error probability of strong users.
Indeed, for one particular user $U_A$, if more power is allocated to the weaker users (with higher SIC priorities), although $U_A$ can decode their messages more accurately, the accuracy of decoding its own message may decrease due to the less allocated power.
To show this, we consider a two-user SISO-NOMA system, where $4$-QAM is used for strong user $U_2$ and BPSK is used for weak user $U_1$. It can be seen from Fig. \ref{power_ser} that at high SNR, ${\rm SER_{22}}$ (for $U_2$ decoding $S_2$) first decreases and then increases as $P_1$ (the power allocated to $U_1$) increases. However, at low SNR, increasing $P_1$ has no positive effect on ${\rm SER_{22}}$, since the effect of SIC error propagation is less dominant in the case.

According to the above example, there are two issues worth further investigation: 1) the explicit relationship between ${\rm SER_{22}}$ and ${\rm SER_{21}}$ (for $U_2$ decoding $S_1$) under given practical modulation and finite coding length and 2) a quantitative definition of SIC residue or SIC-based achievable rate. Both are important for practical NOMA/MIMO-NOMA system designs from the perspective of the SIC-stability.

\subsection{Massive/Millimeter-Wave MIMO-NOMA}\label{future}
The application of NOMA in massive MIMO systems, as an enhanced version of MIMO-NOMA, especially in the millimeter-wave (mmWave) band, is highly anticipated. Due to some features of massive MIMO, including massive antenna arrays, spatially correlated or beamspace sparse channels, and possible hybrid beamforming (analog/baseband domain) structures \cite{Bichai_Spectrum}, existing MIMO-NOMA designs may be inapplicable or need to be modified. Specifically, as the antenna number increases, the overhead of channel training and feedback becomes prohibitive. This forces a reconsideration of the assumption of perfect CSI at the BS in existing MIMO-NOMA works. Consequently, it is desirable to design statistical CSI aided/based MIMO-NOMA schemes or exploit the NOMA characteristics to relax the demand on CSI completeness or quality. New formulations and solutions need to be discovered by carefully exploiting massive MIMO and mmWave features and the properties of NOMA. In the following, three main aspects for massive/mmWave MIMO-NOMA are discussed.

\subsubsection{Channel Estimation and Limited Feedback for Massive MIMO-NOMA}
Plenty of spatial degrees of freedom provided by massive antennas at the BS are viewed as the key performance booster for massive MIMO. However, due to the large antenna number and user number, the channel estimation and feedback of beamforming/power allocation/user clustering information will bring huge system overhead.
Compressive sensing (CS) based channel estimation \cite{Bichai_Spectrum} was shown to be efficient for massive/mmWave MIMO. Thus, CS-based channel estimation for massive MIMO-NOMA is a promising research direction. Either the random CS with omnidirectional training pilots for massive MIMO or the hierarchical codebook based beam alignment for mmWave MIMO should be redesigned to embody NOMA features such as the near-far-user effect. Especially when the mmWave band is used, most existing beam alignment algorithms can only be implemented for one user at a time, which needs to be improved to accommodate the large user number in massive MIMO-NOMA.

Limited feedback for massive/mmWave MIMO-NOMA has been studied but with limited results. In \cite{Ding_Massive}, the use of one-bit feedback achieves the same diversity order as that with perfect user ordering for systems with two-users clusters. In \cite{Ding_millimeter}, the feedback of distance information were compared with one bit feedback of channel quality where the latter was shown to be better with carefully designed quantization threshold. However, since no instantaneous CSI is used for beamforming and power allocation in the above two works, only the feedback of the user ordering information is considered. The more complicate limited feedback of large dimension CSI can potentially provide better performance and deserves further studies.

\subsubsection{Hybrid Massive MIMO-NOMA}\label{hybrid massive}
Due to the small antenna spacing, massive MIMO channels are usually rank deficient and spatially correlated. Another method to circumvent the difficulty of full instantaneous CSI acquisition in massive MIMO-NOMA is to jointly exploit the statistical channel correlation and dimension reduced instantaneous CSI. In \cite{Ding_Massive}, Z. Ding et al. provided a scheme for the multiple-antenna user case. With users' channel correlation at the BS, the users are firstly divided into several groups, where users in each group have the same correlation. Then correlation-based beamforming is used to suppress the inter-group interference and assign $\tilde{M}$ beams to each group. Since the BS serves more than $\tilde{M}$ users in each group, users within each group are further divided into several clusters, where users in each cluster share one of the $\tilde{M}$ effective transmit antennas and ZFBF are used at the user side to avoid inter-cluster interference.

\begin{figure}[t]
\centering
\includegraphics[scale=0.6]{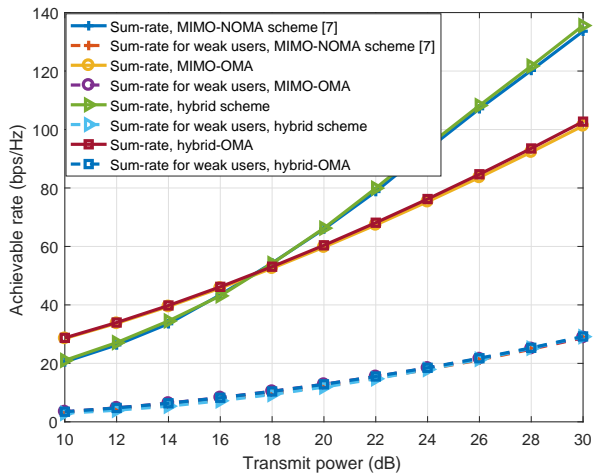}
\captionsetup{margin=5pt,font=small}
\caption{Comparison among the MIMO-NOMA scheme \cite{Kim_Non}, the hybrid scheme, and their OMA counterparts. $M=32$, $K=64$. Path loss exponent: $2$, cell radius: $100$m. The transmitted power is normalized by the path loss of the nearest user ($10$m far away from the BS). The receiver noise power is normalized to 1.}\label{comp_massive}
\end{figure}

For the more practical case of single-antenna users, ZFBF at the user side is infeasible. Instead,
beamforming based on the instantaneous CSI of the reduced $\tilde{M}$ dimension channel can be used to suppress inter-cluster interference. The performance of this hybrid scheme (based on both statistical and reduced instantaneous CSI) is compared with that of the MIMO-NOMA scheme in \cite{Kim_Non}, and their OMA counterparts in Fig. \ref{comp_massive} for a system with 32 BS antennas and 64 single-antenna users. For simplicity, the discrete fourier transform (DFT) matrix is used as the channel eigenvector matrix \cite{Zhang_statistical} with 32 discrete channel directions. Assume that among all the users, 32 users have channel directions $1$ through $16$ while the other 32 users have channel directions $17$ through $32$. They are divided into 2 groups which reduces the training pilot length by half. The figure shows that this hybrid scheme achieves the same performance with the MIMO-NOMA scheme but with much lower training overhead.

Note that statistical CSI based perfect user grouping and instantaneous CSI based user clustering are considered in the above simulation.
For massive MIMO-NOMA systems, practical low complexity approaches for statistical CSI based dynamic grouping and user clustering are of significant importance.
Existing ideas on statistical CSI-based massive MIMO designs can be explored and remodeled to embody the near-far effect in NOMA. The above hybrid beamforming naturally adapts to the hybrid structure where the statistical CSI-based beamforming can be implemented at the analog domain with little performance loss and the reduced instantaneous CSI based beamforming can be implemented in the low dimension baseband domain.

\subsubsection{Statistical Massive MIMO-NOMA}\label{Statistical massive MIMO-NOMA}
\begin{figure}[t]
\centering
\includegraphics[scale=0.6]{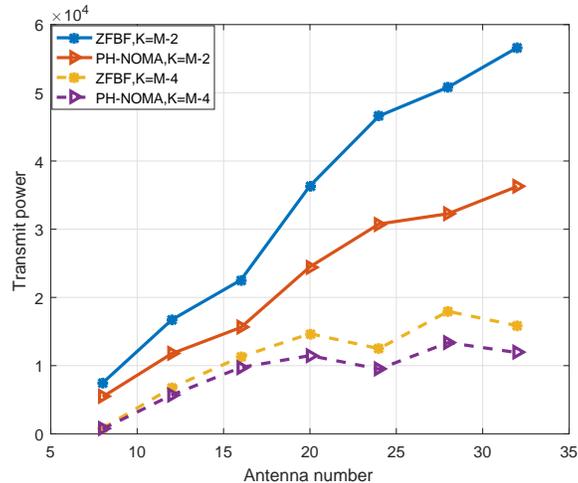}
\captionsetup{margin=5pt,font=small}
\caption{The effect of NOMA on MIMO performance due to more antennas at the BS. Path loss exponent: $2$, cell radius: $100$m. The nearest user is $10$m far away from the BS. The receiver noise power is normalized to 1.}\label{MgeK}
\end{figure}

In existing works, the performance improvement for MIMO-NOMA due to additional antenna number has been evaluated for the full instantaneous CSI case. It can be shown from the performance comparison between linear ZFBF and PH-NOMA \cite{zhiyong_beamforming} in Fig.~\ref{MgeK}
that PH-NOMA needs less transmit power to guarantee users' rate/SINR constraints. This advantage becomes larger as $M$ increases or $M-K$ decreases, which shows the positive effect of NOMA on massive MIMO systems especially when $K$ is large. However, for massive MIMO systems under dense/high mobility deployments, where the channel coherence time is short, the overhead for the reduced instantaneous CSI estimation as mentioned in Section \ref{hybrid massive} can still be unaffordable. Thus, beamforming design  based on statistical CSI only \cite{Zhang_statistical}, e.g., channel correlation, channel direction, or coarse user location, under the NOMA principle is desirable. Compared with the instantaneous CSI counterpart studied in \cite{zhiyong_beamforming}, \cite{Minorization_Hanif,Zhiyong_quasi}, the statistical Massive MIMO-NOMA design tends to be more challenging due to the complicated relationship between optimization metric, e.g. sum-rate, and beamforming vectors\cite{Zhang_statistical}.

\section{Conclusion}
This paper provides an overview on current MIMO-NOMA developments and discussions on some future aspects of MIMO-NOMA. First, basic concepts and key problems of MIMO-NOMA were illustrated. Then, existing works on MIMO-NOMA were reviewed from the aspect of single-cluster and multi-cluster MIMO-NOMA, respectively. By focusing on the problem formulation and key principles of beamforming, user clustering, and power allocation, the contributions and limitations of existing works were discovered. In addition, the issue of the SIC-stability was emphasized for NOMA/MIMO-NOMA system designs with practical modulation and coding schemes. Finally, the combination of NOMA and massive MIMO/millimeter wave MIMO were discussed and the challenges and future research directions were given.

\section{Acknowledgments}
This work was partly supported by the National Natural Science Foundation of China under Grants 61720106003, 61571107 and 61711540305, the Research Project of Jiangsu Province under Grants BE2015156 and BK20160069, and the Scientific Research Foundation of Graduate School of Southeast University.



\begin{thebibliography}{99}
\bibitem{Dai_mag}L. Dai, B. Wang, Y. Yuan, S. Han, C. l. I and Z. Wang, ``Non-orthogonal multiple access for 5G: solutions, challenges, opportunities, and future research trends," \emph{IEEE Commun. Mag.}, vol. 53, no. 9, pp. 74-81, Sep. 2015.
\bibitem{Zhiyong_quasi}Z. Chen, Z. Ding, X. Dai and G. K. Karagiannidis, ``On the application of quasi-degradation to MISO-NOMA downlink," \emph{IEEE Trans. Signal Process.}, vol. 64, no. 23, pp. 6174-6189, Dec. 2016.
\bibitem{Jianyue_Optimal}J. Zhu, J. Wang, Y. Huang, S. He, X. You and L. Yang, ``On optimal power allocation for downlink non-orthogonal multiple access systems," \emph{IEEE J. Sel. Areas Commun.}, vol. 35, no. 12, pp. 2744-2757, Dec. 2017.
\bibitem{Wang_SPL}J. Wang, Q. Peng, Y. Huang, H. M. Wang and X. You, ``Convexity of weighted sum rate maximization in NOMA systems," \emph{IEEE Signal Process. Lett.}, vol. 24, no. 9, pp. 1323-1327, Sep. 2017.
\bibitem{Minorization_Hanif}M. F. Hanif, Z. Ding, T. Ratnarajah and G. K. Karagiannidis, ``A minorization-maximization method for optimizing sum rate in the downlink of non-orthogonal multiple access systems," \emph{IEEE Trans. Signal Process.}, vol. 64, no. 1, pp. 76-88, Jan. 2016.
\bibitem{Ding_Massive}Z. Ding and H. V. Poor, ``Design of massive-MIMO-NOMA with limited feedback," \emph{IEEE Signal Process. Lett.}, vol. 23, no. 5, pp. 629-633, May 2016.
\bibitem{Kim_Non}B. Kimy, S. Lim, H. Kim, S. Suh, J. Kwun, S. Choi, C. Lee, S. Lee and D. Hong, ``Non-orthogonal multiple access in a downlink multiuser beamforming system," in \emph{Proc. IEEE Military Commun. Conf. (MILCOM)}, San Diego, CA, Nov. 2013.
\bibitem{Kim_Design}J. Kim, J. Koh, J. Kang, K. Lee and J. Kang, ``Design of user clustering and precoding for downlink non-orthogonal multiple access (NOMA)," in \emph{Proc. IEEE Military Commun. Conf. (MILCOM)}, Tampa, FL, Oct. 2015.
\bibitem{Ali_Non}S. Ali, E. Hossain and D. I. Kim, ``Non-orthogonal multiple access (NOMA) for downlink multiuser MIMO systems: user clustering, beamforming, and power allocation," \emph{IEEE Access}, vol. 5, pp. 565-577, 2017.
\bibitem{Choi_Minimum}J. Choi, ``Minimum power multicast beamforming with superposition coding for multiresolution broadcast and application to NOMA systems," \emph{IEEE Trans. Commun.}, vol. 63, no. 3, pp. 791-800, Mar. 2015.
\bibitem{zhiyong_beamforming}Z. Chen, Z. Ding and X. Dai, ``Beamforming for combating inter-cluster and intra-cluster interference in hybrid NOMA systems," \emph{IEEE Access}, vol. 4, pp. 4452-4463, 2016.
\bibitem{Shanzhi_Pattern}S. Chen, B. Ren, Q. Gao, S. Kang, S. Sun and K. Niu, ``Pattern division multiple access—a novel nonorthogonal multiple access for fifth-generation radio networks," \emph{IEEE Trans. Veh. Technol.}, vol. 66, no. 4, pp. 3185-3196, Apr. 2017.
\bibitem{Bichai_Spectrum}B. Wang, L. Dai, Z. Wang, N. Ge and S. Zhou, ``Spectrum and energy efficient beamspace MIMO-NOMA for millimeter-wave communications using lens antenna array," \emph{IEEE J. Sel. Areas Commun.}, vol. 35, no. 10, pp. 2370-2382, Oct. 2017.
\bibitem{Ding_millimeter}Z. Ding, P. Fan and H. V. Poor, ``Random beamforming in millimeter-wave NOMA networks," \emph{IEEE Access}, vol. 5, pp. 7667-7681, 2017.
\bibitem{Zhang_statistical}C. Zhang, Y. Huang, Y. Jing, S. Jin and L. Yang, ``Sum-rate analysis for massive MIMO downlink with joint statistical beamforming and user scheduling," \emph{IEEE Trans. Wireless Commun.}, vol. 16, no. 4, pp. 2181-2194, Apr. 2017.


\end{thebibliography}
\end{document}